
\magnification=1200
\hsize=15truecm
\vsize=23truecm
\baselineskip 20 truept
\voffset=-0.5truecm
\parindent=1cm
\overfullrule=0pt

\centerline
{\bf HIGH ENERGY ASYMPTOTICS OF MULTI--COLOUR QCD}

\centerline
{\bf AND EXACTLY SOLVABLE LATTICE MODELS}

\vskip 1truecm

\centerline{\bf L.N. Lipatov}

\vskip 0.5truecm

\centerline{\sl St. Petersburg Nuclear Physics Institute}

\smallskip

\centerline{\sl Gatchina, 188350, Russia}

\vskip 1truecm

\centerline{\sl Istituto Nazionale di Fisica Nucleare}

\smallskip

\centerline{\sl Sezione di Padova}

\centerline{\sl Via Marzolo 8, 35131 Padova (Italy)}

\vskip 1.5truecm

\centerline{\bf Abstract}

\vskip 0.5truecm

\noindent
The quantum inverse scattering method is applied to the solution
of equations for wave functions of compound states of $n$ reggeized
gluons in the multicolour QCD in a generalized leading logarithmic
approximation.

\vskip 4truecm

\noindent
DFPD/93/TH/70\hfill October 1993

\vfill\eject

In the leading logarithmic approximation (LLA), where $g^2
\ell n S\sim 1$ ($g$ is QCD coupling constant), the asymptotics
of amplitudes for the colourless particle scattering at high
energies $\sqrt{S}$ is governed by the Feynman diagrams with
two reggeized gluons in the crossing channel $^{[1]}$. The
intercept of the Pomeron trajectory turns out to be bigger
than unity in disagreement with the Froissart bound $^{[2]}$,
which is a consequence of the violation of $S$--channel
unitarity constraints in LLA. One of the possible method of unitarizing
the LLA results is based on the effective field theory valid for
the multi--Regge kinematics of produced gluons $^{[3]}$. Another
(approximate) method corresponds to a generalized LLA, in which one
calculates the contributions of the Feynman diagrams with an arbitrary
but conserved number $n$ of reggeized gluons in the crossing channel
$^{[4]}$. In our previous work it was shown that the corresponding
Bethe--Salpeter equations for the $n$ gluon t--channel partial
waves $f_\omega, \omega=j-1$, are significantly simplified in the multi--colour
limit $N\rightarrow\infty$ $^{[5]}$. In the impact parameter
representation the wave functions $f_w(\vec\rho_1, \vec\rho_2....
\vec\rho_n; \vec\rho_0)$ for compound states of the reggeized
gluons with two--dimensional transverse coordinates
$(\vec\rho_1, \vec\rho_2...\vec\rho_n)$ have the property of the
holomorphic factorization:

$$
f_\omega(\vec\rho_1, \vec\rho_2...\vec\rho_n; \vec\rho_0)=
\sum_r f^r (\rho_1, \rho_2,...\rho_n;\rho_0)
\tilde f^r (\rho^*_1, \rho^*_2,...f^*_n; \rho^*_0),\eqno(1)
$$

\noindent
where $f^r(\tilde f^r)$ are holomorphic (antiholomorphic) functions
of the complex coordinates $\rho_i, \rho^*_i$ of gluons in the
impact parameter space. The position $\omega$ of singularities of
t--channel partial waves is expressed in terms of energies
$\epsilon$, $\tilde\epsilon$ for corresponding holomorphic and
antiholomorphic subsystems:

$$
\omega=-{g^2\over 16\pi^2}\ N(\epsilon+\tilde\epsilon),\quad
\epsilon f^r = {\cal H} f^r, \quad \tilde\epsilon f^r =
{\cal H}^* f^r,\eqno(2)
$$

\noindent
where the total holomorphic Hamiltonian is

$$
{\cal H} = \sum^n_{i=1} H_{i,i+1}. \eqno(3)
$$

\noindent
The pair Hamiltonians $H_{i,k}$ can be written in two equivalent
forms

$$\eqalign{
H_{ik} & = P^{-1}_i \ell n (\rho_{ik}) P_i + P^{-1}_k \ell n
(\rho_{ik}) P_k + \ell n (P_i P_k) - 2\psi(1) = \cr
& = 2\ell n (\rho_{ik}) + \rho_{ik} \ell n (P_i P_k) \rho_{ik}^{-1}
-2\psi(1),\cr}\eqno(4)
$$

\noindent
where $\rho_{ik} = \rho_i -\rho_k, P_k = i{\partial\over
\partial\rho_k}, \psi(1)=-\gamma,\gamma$ is the Euler constant.
{}From representations (4) one can verify that the operator
${\cal H}^T$, conjugated to ${\cal H}$, can be constructed from
${\cal H}$ with the use of two different similarity transformations:

$$
H^T = P_n^{-1}...P^{-1}_1 HP_1...P_n = \rho_{12} \rho_{23}...
\rho_{n1} H \rho^{-1}_{12} \rho_{23}^{-1}...
\rho^{-1}_{n1}.\eqno(5)
$$

\noindent
Therefore there is the following differential operator, commuting
with Hamiltonian (3):

$$
{\cal A} = \rho_{12} \rho_{23}...\rho_{n1} P_1 P_2...P_n,\
[{\cal A}, {\cal H}] =0\eqno(6)
$$

\noindent
and, to find a solution of Schr\"odinger's eqs. (2), it is enough
to find all solutions of the differential equation $^{[5]}$:

$$
{\cal A} f = \lambda f.\eqno(7)
$$

\noindent
In this paper we shall show that there is a set of n--th order
mutually commuting
differential operators, including ${\cal A}$ (6) and being a
solutions of the Yang--Baxter equations for a monodromy matrix.

To begin with, let us introduce the following $2\times 2$
monodromy matrix

$$
T(\Theta) =
\left(\matrix{
\Theta  +\rho_1\partial_1, & \partial_1\cr
 -\rho^2_1\partial_1, & \Theta-\rho_1\partial_1\cr}\right)\
\left(\matrix{
\Theta  +\rho_2\partial_2, & \partial_2\cr
-\rho^2_2\partial_2,&\Theta-\rho_2\partial_2\cr}\right)\ldots
$$

$$
\ldots
\left(\matrix{
\Theta +\rho_n\partial_n, & \partial_n\cr
 -\rho^2_n\partial_n, & \Theta-\rho_n\partial_n\cr}\right)\eqno(8)
$$

\noindent
where $\Theta$ is the spectral parameter, $\partial_k = {\partial \over
\partial \rho_k}$.

\noindent
One can easily verify that its factors

$$
L(\Theta) =
\left(\matrix{
\Theta + \rho\partial, &\partial\cr
- \rho^2\partial, &\Theta-\rho\partial\cr}\right)\eqno(9)
$$

\noindent
satisfy the following commutation relations

$$
L(v)_{i_2 k_2} L(u)_{i_1 k_1}
L^{1/2}_{k_1 k_2, j_1 j_2} (u-v) =
L^{1/2}_{i_1 i_2, k_1 k_2} (u-v) L(u)_{k_2 j_2}
L(v)_{k_1 j_1},\eqno(10)
$$

\noindent
where the numerical matrix $L^{1/2}_{k_1 k_2;j_1 j_2}(u)$ is the
lattice L--operator for the XXX Baxter model with the following
nonvanishing matrix elements:

$$\eqalign{
L^{1/2}_{11;11}(u) & = L^{1/2}_{22;22} (u)= u+1,\cr
L^{1/2}_{12;12}(u) & = L^{1/2}_{21;21} = u,\cr
L^{1/2}_{12;21}(u) & = L^{1/2}_{21;12} = 1.\cr}
\eqno(11)
$$

\noindent
The matrix $L^{1/2}$ satisfies the three--linear Yang--Baxter
equations $^{[6]}$. As a consequence of eq. (10) we obtain the
commutation relations for the matrix elements of $T(\Theta)$
(8):

$$
T(v)_{i_2 k_2} T(u)_{i_1 k_1} L^{1/2}_{k_1 k_2, j_1 j_2} (u-v)=
L^{1/2}_{i_1 i_2, k_1 k_2} (u-v) T(u)_{k_2 j_2} T(v)_{k_1 j_1}.
\eqno(12)
$$

\noindent
{}From these equations one can conclude in particular, that the operators

$$
t(\Theta) = T (\Theta)_{11} + T (\Theta)_{22} \eqno(13)
$$

\noindent
commute with each others

$$
t(v) t(u) = t(u) t(v). \eqno(14)
$$

\noindent
Therefore, if we expand $t(\Theta)$ in the series over $\Theta$

$$
t(\theta)= Q_n + \theta Q_{n-1}+...+ \theta^h, \eqno(15)
$$

\noindent
then the coefficients $Q_k$, being the k--th order differential operators:

$$
Q_k =
\sum_{ \{i_1,i_2,...i_k\}\atop i_1<i_2<...<i_k}
\rho_{i_1 i_2} \rho_{i_2 i_3...} \rho_{i_k
i_1} \partial_{i_1} \partial_{i_2...} \partial_{ik},
\eqno(16)
$$

\noindent
commute

$$
[Q_k, Q_{k'}] = 0 \eqno(17)
$$

\noindent
The operator ${\cal A}$ (7) is proportional to $t(0)$:

$$
{\cal A} = i^n Q_n \eqno(18)
$$

\noindent
Another operator, which also commutes with ${\cal H}$ (3) is

$$
Q_2 = - \sum_{i<k} \rho^2_{ik} \partial_i \partial_k. \eqno(19)
$$

\noindent
$Q_2$ is the Kazimir operator $(\sum_i \vec M_i)^2$ of the M\"obious group
of transformations:
$\rho \rightarrow {a\rho + b \over c \rho + d}$, concerning which eqs. (2)
are invariant $^{[2,5]}$.

\noindent
It is plausible that all operators $Q_k$ commute with ${\cal H}$. Moreover
it is natural to expect that ${\cal H}$ is a function of operators $Q_k$ and
therefore to find its eigenvalues it is enough to calculate eigenvalues
of $Q_k$. Eigenvalues $\lambda$ and eigenfunctions of eq. (7) can be
obtained by using the algebraic Bethe ansatz $^{[6]}$. In the future
publications we shall use the quantum inverse
scattering method, developed by L. Faddeev and his
collaborators $^{[6]}$, to
solve eqs. (2), (7), starting from commutation relations (12).

\noindent
The remarkable mathematical properties of the equations for QCD scattering
amplitudes at high energies could be explained from the fact, that the
Yang--Mills theory  is a low energy limit of the super--string model,
having a very rich symmetry group.

\vskip 1truecm
\noindent
The author is thankful to the Dipartimento di Fisica dell'Universit\`a
di Padova for the hospitality and to Prof.
A. Bassetto and Dr. M. Matone for helpful discussions.

\vskip 1.5truecm

\centerline{\bf References}

\vskip 0.3truecm

\noindent
1. E.A. Kuraev, L.N. Lipatov and V.S. Fadin, Sov. Phys. JETP \underbar{44}
(1976) 443; \underbar{45} (1977) 199;

\smallskip

\noindent
2. L.N.Lipatov, Sov. Phys. JETP \underbar{63} (5) (1986) 904;

\smallskip

\noindent
3. L.N. Lipatov, Nucl. Phys. \underbar{B365} (1991) 641;

\smallskip

\noindent
4. J. Bartels, Nucl. Phys. \underbar{B175} (1980) 365;
J. Kwiecinski and M. Praszalowicz, Phys. Lett. \underbar{B94} (1980) 413;

\smallskip

\noindent
5. L.N. Lipatov, Physics Lett. \underbar{B309} (1993) 394;

\smallskip

\noindent
6. L.D. Faddeev, Soviet Sci. Reviews, Contemporary Math. Phys.,
\underbar{C1} (1980) 107.

\bye